\documentclass[a4paper,aps,prd,twocolumn,10pt,superscriptaddress,nofootinbib,showpacs]{revtex4-1}
\usepackage{graphicx,amsmath,amsfonts,amssymb,revsymb,dcolumn,epsfig,bm,xspace}
\usepackage[hyperfootnotes=false]{hyperref}
\usepackage[latin1]{inputenc}
\usepackage{color}
\usepackage{accents}

\newcommand{\lie}{\pounds}

\newcommand{\la}{\langle}
\newcommand{\ra}{\rangle}

\newcommand{\M}{M}
\newcommand{\Mback}{M_{bg}}
\newcommand{\SM}{\Sigma}
\newcommand{\bSM}{\wbackdec{\SM}}
\newcommand{\gv}{A}
\newcommand{\gvpa}{\mathcal{A}}
\newcommand{\gvpe}{\mathsf{A}}
\newcommand{\Weyl}{W}
\newcommand{\WeylE}{E}
\newcommand{\overbar}[1]{\mkern 1.8mu\overline{\mkern-1.8mu#1\mkern-1.8mu}\mkern 1.8mu}
\newcommand{\backdec}[1]{\bar{#1}}
\newcommand{\wbackdec}[1]{\overbar{#1}}
\newcommand{\g}{g}
\newcommand{\dg}{{\delta\g}}
\newcommand{\bg}{{\backdec{g}}}
\newcommand{\gback}{q}
\newcommand{\n}{v}
\newcommand{\bn}{{\backdec{\n}}}
\newcommand{\dn}{{\delta\n}}
\newcommand{\sn}{\mathsf{v}}
\newcommand{\sns}{\mathcal{V}}
\newcommand{\snv}{\mathsf{V}}
\newcommand{\h}{\gamma}
\newcommand{\bh}{\backdec{\h}}

\newcommand{\hp}[1]{\h\left[#1\right]}
\newcommand{\bhp}[1]{\bh\left[#1\right]}
\newcommand{\cd}{\nabla}
\newcommand{\bcd}{\wbackdec{\nabla}}
\newcommand{\scd}{D}
\newcommand{\bscd}{\wbackdec{\scd}}
\newcommand{\scp}{\parallel}
\newcommand{\blap}{\wbackdec{\scd}^2}
\newcommand{\blapK}{\blap_{\K}}
\newcommand{\blapR}{\blap_{\bSR}{}}
\newcommand{\bilap}{\wbackdec{\scd}^{-2}}
\newcommand{\bilapK}{\bilap_{\K}}
\newcommand{\bilapR}{\bilap_{\bSR}{}}

\newcommand{\kp}{\kappa}

\newcommand{\A}{\phi}
\newcommand{\AI}{\Phi}
\newcommand{\AIt}{\mathcal{J}}
\newcommand{\dAIt}{\delta\AIt}
\newcommand{\B}{B}
\newcommand{\C}{C}
\newcommand{\CS}{\psi}
\newcommand{\CSI}{\Psi}
\newcommand{\CSIt}{\mathcal{T}}
\newcommand{\dCSIt}{\delta\CSIt}

\newcommand{\CTD}{W}
\newcommand{\BS}{\mathcal{\B}}
\newcommand{\BV}{\mathtt{\B}}
\newcommand{\CSD}{\mathcal{E}}
\newcommand{\CV}{\mathtt{F}}
\newcommand{\MP}{\mathcal{P}}
\newcommand{\bMP}{\wbackdec{\mathcal{P}}}

\newcommand{\ggt}{B}
\newcommand{\EC}{\mathcal{K}}
\newcommand{\bEC}{\wbackdec{\EC}}

\newcommand{\EX}{\Theta}
\newcommand{\gEX}{Z}
\newcommand{\dgEX}{\mathcal{Z}}
\newcommand{\dEX}{{\delta\EX}}

\newcommand{\EXI}{\Xi}
\newcommand{\EXIt}{\mathcal{U}}
\newcommand{\dEXIt}{\delta\EXIt}

\newcommand{\bEX}{\wbackdec{\EX}}
\newcommand{\SH}{\sigma}
\newcommand{\dSH}{\delta\sigma}
\newcommand{\dSHs}{\mathcal{S}}
\newcommand{\dSHv}{\mathsf{S}}

\newcommand{\SR}{\mathcal{R}}
\newcommand{\SRtl}{r}

\newcommand{\bSR}{\wbackdec{\SR}}
\newcommand{\K}{K}
\newcommand{\bK}{\wbackdec{\K}}
\newcommand{\ac}{a}
\newcommand{\Ac}{\alpha}
\newcommand{\dAc}{\delta\Ac}
\newcommand{\bac}{\backdec{\ac}}
\newcommand{\dac}{{\delta\ac}}

\newcommand{\pureac}{\ring{\ac}}
\newcommand{\puredac}{\delta\ring{\ac}}
\newcommand{\pureSH}{\ring{\SH}}
\newcommand{\puredSH}{\delta\ring{\SH}}
\newcommand{\puredEX}{\delta\ring{\EX}}
\newcommand{\pureSRtl}{\ring{\SRtl}}
\newcommand{\puredSRtl}{\delta\ring{\SRtl}}

\newcommand{\pureged}{\ring{\ged}}
\newcommand{\pureded}{\delta \ring{\ed}}
\newcommand{\puregp}{ \ring{\gp}}
\newcommand{\puredp}{\delta \ring{\p}}
\newcommand{\puregEX}{\ring{\gEX}}
\newcommand{\puredgEX}{\delta \ring{\dgEX}}
\newcommand{\purednm}{\delta \ring{\nm}}
\newcommand{\purenm}{ \ring{\nm}}
\newcommand{\pureed}{\ring{\ed}}
\newcommand{\purep}{\ring{\p}}
\newcommand{\pureAP}{\ring{\AP}}
\newcommand{\pureEX}{\ring{\EX}}

\newcommand{\T}{T}

\newcommand{\mf}{\varphi}
\newcommand{\dmf}{\delta\mf}
\newcommand{\mfIt}{\mathcal{M}}
\newcommand{\dmfIt}{\delta\mfIt}
\newcommand{\nm}{q}

\newcommand{\snm}{\mathsf{u}}
\newcommand{\snms}{\mathcal{U}}
\newcommand{\snmv}{\mathsf{U}}
\newcommand{\bmf}{\backdec{\mf}}

\newcommand{\AP}{\Pi}

\newcommand{\p}{p}
\newcommand{\gp}{Y}
\newcommand{\dpp}{\delta\p}

\newcommand{\ed}{\rho}
\newcommand{\ged}{X}

\newcommand{\ded}{\delta\ed}

\newcommand{\bp}{\backdec{p}}
\newcommand{\bed}{\backdec{\ed}}

\begin{document}
\title{Covariant Bardeen Perturbation Formalism}

\author{S.~D.~P.~Vitenti} \email{dias@iap.fr}
\affiliation{CAPES Foundation, Ministry of Education of Brazil, Bras\'{\i}lia -- DF 70040-020, Brazil}
\affiliation{${\mathcal G}\mathbb{R}\varepsilon\mathbb{C}{\mathcal O}$ -- Institut d'Astrophysique de Paris, UMR7095 CNRS, Universit\'e Pierre \& Marie Curie, 98 bis boulevard Arago, 75014 Paris, France}

\author{F.~T.~Falciano} \email{ftovar@cbpf.br}
\author{N.~Pinto-Neto} \email{nelson.pinto@pq.cnpq.br}

\affiliation{Centro Brasileiro de Pesquisas F\'{\i}sicas,
Rua Dr.\ Xavier Sigaud 150 \\
22290-180, Rio de Janeiro -- RJ, Brasil}

\date{\today}

\begin{abstract}
In a previous work we obtained a set of necessary conditions for the linear approximation in cosmology. Here we discuss the relations of this approach with the so called covariant perturbations. It is often argued in the literature that one of the main advantages of the covariant approach to describe cosmological perturbations is that the Bardeen formalism is coordinate dependent. In this paper we will reformulate the Bardeen approach in a completely covariant manner. For that, we introduce the notion of pure and mixed tensors, which yields an adequate language to treat both perturbative approaches in a common framework. We then stress that in the referred covariant approach one necessarily introduces an additional hyper-surface choice to the problem. Using our mixed and pure tensors approach, we were able to construct a one-to-one map relating the usual gauge dependence of the Bardeen formalism with the hyper-surface dependence inherent to the covariant approach. Finally, through the use of this map, we define full non-linear tensors that at first order correspond to the three known gauge invariant variables $\AI$, $\CSI$ and $\EXI$, which are simultaneously foliation and gauge invariant. We then stress that the use of the proposed mixed tensors allows one to construct simultaneously gauge and hyper-surface invariant variables at any order.
\end{abstract}


\maketitle

\section{Introduction}

The cosmological perturbations formalism provide an important tool to study the evolution of inhomogeneities in the standard cosmological model. However, since its initial development~\cite{Lifshitz1946,Lifshitz1963}, the gauge-problem has raised several issues concerning the physical interpretation of such perturbations. After a quite long discussion, it appeared in the literature two main approaches to deal with such difficulties. In a pioneering paper \cite{Hawking1966} Hawking introduced the so called covariant approach, which was later used by many authors \cite{Ellis1989, Ellis1990, Dunsby1992,
Bruni1992, Bruni1992a, Novello1995, Novello1995a, Novello1996, Dunsby1997, Tsagas1997, Challinor1998, MacCallum2000, Novello2000, Langlois2005, Langlois2005a, Langlois2006, Langlois2007,Tsagas2008} to define gauge invariant (GI) variables associated with the gradients of background scalar quantities and the Weyl tensor. The second approach was developed by Gerlach and Sengupta \cite{Gerlach1978} and then by Bardeen \cite{Bardeen1980} in the cosmological scenario. In the latter paper, Bardeen constructed GI variables as combinations of the metric perturbations within a specific coordinate system. Even though Bardeen's approach may seem straightforward, its implementation rapidly becomes very involved, and in a sense continues to be coordinate dependent since one has to start with gauge dependent equations and variables and then combine them in a gauge invariant form.

Generally, metric perturbations are defined through the specification of a background metric written in a specific coordinate system, and then defining its perturbations as differences of the physical metric tensor from this fiducial background metric. In this scenario, the gauge freedom that appears in the perturbations are associated with the freedom in the different possible ways to map the physical space-time with respect to the background manifold used as a reference  (see for instance \cite{Mukhanov1992}).

The physical meaning of Bardeen's GI variables has been discussed in~\cite{Bardeen1980} by analyzing them in different gauges. Complementarily to this work and, as an effort to make a contact between the two formalisms, in \citet{Bruni1992} the authors investigated their meaning by using the covariant scenario. Their approach was to calculate the covariant perturbations in a specific coordinate system and compare them with Bardeen's variables.

Apart from the GI variables, the meaning of the gauge transformations themselves has also been investigated by Stewart and Walker in \cite{Stewart1974,Stewart1990}. Among other results, in these papers the authors address the tensorial decomposition into scalar, vector and tensor variables. Their approach consists in considering a smooth family of space-times parametrized by some parameter $\epsilon$, and the perturbation as the first order expansion of the metric with respect to this parameter. This method has also been used by Wald \cite{Wald1984} to give a formal description of a generic metric perturbation.

The purpose of this work is to understand more deeply the relation between the two above mentioned approaches to the theory of cosmological perturbations, and to present some new improvements in these formalisms. First of all, we present the Bardeen approach in a complete covariant way. With this aim, we introduce a new classification of tensors in terms of the fundamental objects used in their definitions. We call mixed tensors the elements of the set of all geometrical objects which necessarily combine background and physical tensors in their definitions, while pure tensors are the elements of its complement. The gauge dependent Bardeen variables can be obtained from these mixed tensors, even if their background values are null or constant because such mixed tensors do not satisfy the Stewart-Walker lemma \cite{Stewart1974,Stewart1990}, only pure tensors do.

On the other hand, the covariant approach defines the cosmological perturbations in terms of pure tensors, using only physical geometrical objects. Hence, even in the case where their background values are null or constant, guaranteeing their gauge invariance, they are intrinsically foliation dependent. As a consequence, while the Bardeen approach has its well known gauge dependence, the covariant formalism contains a dependence on the space-like hyper-surfaces one uses to perform the space-time foliation. Using our new classification, we compared both formalisms. We have shown that, besides their dynamical equivalence, their indeterminacies are complementary. The usual gauge dependence appearing in the Bardeen approach can be one-to-one mapped into the hyper-surface choice dependence.

The natural question that appears is whether there are variables that are simultaneously foliation and gauge invariant. The answer is affirmative, and we have presented a general way to construct full non-linear tensors that at first order are simultaneously foliation and gauge invariant, and how they are related to the electric part of the Weyl tensor, which is also gauge and foliation independent at first order. The definition of these tensors can be made using either pure tensors or mixed tensors, hence they are common to the new covariant Bardeen approach and the usual covariant approach. Their scalar parts correspond exactly to the Bardeen gauge invariant potentials, hence showing that these potentials are a particular combination that at first order are not only gauge invariant, as it is well known, but also foliation independent.

In Sec.~\ref{sec:pert:sp} we will briefly review the metric perturbation formalism, and show how one can define metric perturbations in a completely covariant way. This method is generic and valid for an arbitrary background metric. During this process, we will introduce the concept of pure and mixed tensors, which gives an adequate common language to treat both the already usually covariant~\cite{Hawking1966,Ellis1989} and Bardeen's~\cite{Bardeen1980} approaches in the same framework.

In Sec.~\ref{sec:mixten} we define the kinetic variables for pure and mixed tensors. Additionally, their decomposition are related to the gauge freedom and the choice of the space-time foliation.  By controlling their variations, it is possible to relate these two freedoms, and at first order one can map one into the other directly. As a bonus, our method has shown to be much simpler then others to do calculations in a gauge free manner (without choosing any gauge). For instance, in~\cite{Vitenti2013} (from now on VFP) we have used this method to obtain the perturbed second order Lagrangian for an arbitrary background in a foliation free manner. Most of the perturbative expansions that we shall use here were carefully obtained in VFP, hence we maintain the same notation introduced in Section~II of VFP.

Following this discussion, in Sec.~\ref{sec:gi} we construct nonlinear tensors whose first order perturbations give us the usual GI variables. In particular, the procedure we use to define these tensors is that at first order they should give foliation independent variables. In this way, the Bardeen potentials are special not only for being GI variables, but also for being foliation independent. Finally, in Sec.~\ref{conclusions} we conclude with final remarks.

\section{Perturbations by Means of Pure and Mixed Tensors}
\label{sec:pert:sp}

A gauge theory is one in which the specification of the dynamical variables depends on functions which can be arbitrarily chosen at any instant of time \cite{Henneaux1994}. Any change on this choice is called a gauge transformation. Of course the physical observables of a gauge theory should not depend on these arbitrary functions, hence they should be invariant under gauge transformations (they should be gauge invariant quantities). In electromagnetism, the vector potential depends on the choice of some scalar functions, but the physical electric and magnetic fields do not depend on them, and so they are gauge invariant quantities.

Any generally covariant theory is a gauge theory: the specification of the dynamical variables depends on the coordinate system one is choosing, and this choice is arbitrary. As General Relativity (GR) is a generally covariant theory, it is also a gauge theory, whose gauge group of transformations is the manifold mapping group ($\mathcal{MMG}$). However, this is not a peculiarity of GR as long as any field theory can be put in a general covariant form \cite{Anderson1967,Wald1984}. The difference arises when one wants to establish the symmetry group of GR (sometimes called invariance group, or proper gauge group \cite{Anderson1967,Grishchuk1984}), which is the sub-group of the $\mathcal{MMG}$ which keeps invariant the absolute objects of the theory written in the coordinate system where they assume their simplest form. For instance, in a generally covariant field theory in flat spacetime (Special Relativity), the flat metric is not subjected to any dynamics. It is a given external and absolute object which is present whatever field configuration one constructs. The metric is insensitive to any initial conditions one might give to the physical system. Hence, the symmetry group of any generally covariant field theory in Special Relativity is the Poincar\`e sub-group of the $\mathcal{MMG}$, and the preferred coordinate system in which the flat metric takes its simplest Minkowski form is the Cartesian coordinates.

The peculiarity of GR, which gives rise to its name, relies on the fact that there is no absolute or external object in GR: all quantities, including geometry itself, are subjected to dynamical laws, and any physical change of initial conditions modify the future development of all geometrical objects of the theory. Therefore, the symmetry group of GR coincides with its gauge group, the $\mathcal{MMG}$. As a consequence, there is no preferred coordinate system, or a preferred gauge fixation. In GR, one must rely on gauge invariant quantities, which are the values of tensor fields and/or particle positions with respect to other tensor fields and/or particle positions, including the gravitational field itself (identified with the geometry of space-time). The values of the tensor fields at a point on the manifold are not gauge invariant quantities because such points do not have a physical meaning. As an example, geodesics are not gauge invariant quantities, hence they are not observables, while geodesic deviations are.

Nevertheless, the situation is more involved when one turns to the theory of perturbations in GR, which is the suitable framework when the physical system possesses some underlying approximate symmetry. By this we mean that geometrical objects on the physical manifold change infinitesimally under some restrict group of finite transformations, which is a sub-group of the $\mathcal{MMG}$. In this case, one can define geometrical objects which are exactly invariant under this sub-group of transformations, $\wbackdec{Q}(x)$,\footnote{Here we will omit the indexes for simplicity.} which we call background geometrical objects. In this context, the perturbation of an arbitrary physical tensor $Q$, can be defined as
\begin{equation}
\delta{}Q(x) = Q(x) - \wbackdec{Q}(x),
\end{equation}
where $\wbackdec{Q}$ stands for the background variable, where we assume that the quantity $\delta{}Q(x)$ is small in some sense \cite{Wald1984,Vitenti2012}. Assuming that one has a covariant way to measure the size of the perturbations (see for example~\cite{Green2011}), then, since $\delta{}Q(x)$ is a tensor, any general infinitesimal gauge transformation belonging to the $\mathcal{MMG}$ group in which all objects are transformed will not alter the size of the perturbations. However, there is an arbitrariness on the way the background tensors are introduced in the physical manifold. They are naturally defined in a background manifold $\Mback$ and one should define them in the physical manifold $\M$ through a diffeomorphism. Hence, consider a certain manifold $\Mback$ with a given riemannian metric $\gback_{\mu\nu}$ and other tensor quantities, and a smooth family of diffeomorphisms $\Upsilon_\lambda:\mathrm{I}\times\Mback\rightarrow\M$ from $\Mback$ to the actual physical manifold $\M$ that depends on some parameter $\lambda$ defined in the interval $\mathrm{I} = [0, b]$ for some real number $b$. For each value of $\lambda$, we can define a bijective tensor map from  $\Mback$ to $\M$ as $\Upsilon_\lambda^{*}$.  Through this bijective map, we can define the background metric in $\M$ as $\bg_{\mu\nu} \equiv \Upsilon_0^{*}\gback_{\mu\nu}$.\footnote{For the moment we restrict our analysis to the metric tensor, but the above reasoning can be extended to any other geometrical object defined on $\Mback$. All metrics have signature $(-1,1,1,1)$.} In fact, to make contact with the usual language in cosmological perturbation scenarios, we shall call any tensor brought from $\Mback$ to $\M$ as a background quantity, and any tensor defined strictly on $\M$ as a physical tensor.\footnote{It is often used the term perturbed tensor for these objects. However physical tensor seems more suitable to avoid misunderstanding with the perturbations that are defined as the difference of these objects with background tensors.} In addition, in $\M$ we can also define two distinct covariant derivatives, one for each metric. Thus, we define the covariant derivatives $\cd_\mu$ as the operator compatible with $\g_{\mu\nu}$ and $\bcd_\mu$ compatible with $\bg_{\mu\nu}$.

The main assumption in the perturbation theory is that the difference $$\dg_{\mu\nu} \equiv \g_{\mu\nu} - \bg_{\mu\nu},$$ which we shall call metric perturbation, is small in some sense. For a small $\lambda$, we have that the new background metric reads $$\bg^{(\lambda)}{}_{\mu\nu} = \vartheta^{-1*}_\lambda \bg_{\mu\nu} \approx \bg_{\mu\nu} + \lie_\gv \bg_{\mu\nu},$$ where the infinitesimal vector field $\gv^\mu$, of order $\lambda$, is the tangent field to $p(\lambda) \equiv \Upsilon_\lambda(p)$ in $\M$. Therefore, for small values of $\lambda$, the new metric perturbation reads \begin{equation}
\dg^{(\lambda)}{}_{\mu\nu} = \g_{\mu\nu} - (\bg_{\mu\nu} + \lie_\gv \bg_{\mu\nu}) = \dg_{\mu\nu} - 2\bcd_{(\mu}\gv_{\nu)}.
\label{deltad}
\end{equation}

Note that different values of $\lambda$ define different background metrics on the physical manifold, but this arbitrariness in the introduction of the fiducial background metric must be limited to the above mentioned assumption that one needs $\bcd_{(\mu}\gv_{\nu)}$ to be small in the same sense as $\dg_{\mu\nu}$.\footnote{Note that this is a sufficient but not necessary condition: as mentioned above, if $\gv^\mu$ is a Killing vector of $\bg_{\mu\nu}$ the perturbation remains small for any value of $\lambda$.}

It is worth emphasizing the difference between the above transformation with a general infinitesimal gauge transformation in $\M$ associated with an infinitesimal coordinate transformation. Given $\g_{\mu\nu}$ and $\bg_{\mu\nu}$, which are true tensors in $\M$, such a coordinate transformation generated by an infinitesimal vector field $\ggt^\mu$ changes them as
\begin{align*}
\g_{\mu\nu} \quad &\longrightarrow \quad \g_{\mu\nu} + \lie_\ggt\g_{\mu\nu} \approx \g_{\mu\nu} + \lie_\ggt \bg_{\mu\nu}, \\
\bg_{\mu\nu} \quad &\longrightarrow \quad \bg_{\mu\nu} + \lie_\ggt \bg_{\mu\nu},
\end{align*}
where in the above approximation we neglected second order terms. Clearly, metric perturbations, as well perturbations of any other geometrical object defined in the same way, are invariant under such transformations, i.e.,
\begin{equation}
\widetilde{\dg}_{\mu\nu}\approx (\g_{\mu\nu} + \lie_\ggt \bg_{\mu\nu}) - (\bg_{\mu\nu} + \lie_\ggt \bg_{\mu\nu}) = \dg_{\mu\nu}.
\label{deltac}
\end{equation}

Indeed, $\dg_{\mu\nu}$ being defined as the difference of two tensors it is also a true tensor and as such it is covariant under general coordinate transformations. Note however that $\dg_{\mu\nu}$ is not invariant under the change of diffeomorphism described above [see Eq.~\eqref{deltad} and compare it with Eq.~\eqref{deltac}]. Of course this is true for perturbations of any kind of geometrical objects inasmuch under any diffeomorphism change from $\Upsilon_{0}$ to $\Upsilon_{\lambda}$, all background tensors are transformed, while all physical quantities are kept fixed since they are defined independently of the background manifold $\Mback$ and $\Upsilon_{\lambda}$. For instance, consider an arbitrary background tensor $$\wbackdec{T}_{\nu_1\nu_2\dots\nu_m}^{\quad\mu_1\mu_2\dots\mu_l} \equiv \Upsilon^{*}_0 T^\circ{}_{\nu_1\nu_2\dots\nu_m}^{\quad\mu_1\mu_2\dots\mu_l},$$ and a similar tensor $T_{\nu_1\nu_2\dots\nu_m}^{\quad\mu_1\mu_2\dots\mu_l}$ defined in $\M$. Under a change in the diffeomorphism $\Upsilon_0\ \rightarrow\ \Upsilon_\lambda$ we have
\begin{align*}
\wbackdec{T}_{\nu_1\nu_2\dots\nu_m}^{\quad\mu_1\mu_2\dots\mu_l} \quad &\longrightarrow \quad \wbackdec{T}_{\nu_1\nu_2\dots\nu_m}^{\quad\mu_1\mu_2\dots\mu_l} + \lie_\gv \wbackdec{T}_{\nu_1\nu_2\dots\nu_m}^{\quad\mu_1\mu_2\dots\mu_l}, \\
{T}_{\nu_1\nu_2\dots\nu_m}^{\quad\mu_1\mu_2\dots\mu_l} \quad &\longrightarrow \quad {T}_{\nu_1\nu_2\dots\nu_m}^{\quad\mu_1\mu_2\dots\mu_l}.
\end{align*}

One can then combine the above transformation with a small general coordinate transformation generated by the vector field $\ggt^\mu$, yielding
\begin{align*}
\wbackdec{T}_{\nu_1\nu_2\dots\nu_m}^{\quad\mu_1\mu_2\dots\mu_l} \quad &\longrightarrow \quad \wbackdec{T}_{\nu_1\nu_2\dots\nu_m}^{\quad\mu_1\mu_2\dots\mu_l} + \lie_{\gv+\ggt} \wbackdec{T}_{\nu_1\nu_2\dots\nu_m}^{\quad\mu_1\mu_2\dots\mu_l}, \\
{T}_{\nu_1\nu_2\dots\nu_m}^{\quad\mu_1\mu_2\dots\mu_l} \quad&\longrightarrow \quad {T}_{\nu_1\nu_2\dots\nu_m}^{\quad\mu_1\mu_2\dots\mu_l} + \lie_{\ggt} {T}_{\nu_1\nu_2\dots\nu_m}^{\quad\mu_1\mu_2\dots\mu_l}.
\end{align*}

Hence, the most general way a perturbation can be transformed is given by
\begin{eqnarray}
\delta {T} \equiv T-\wbackdec{T}\quad &\longrightarrow& \quad
\delta {T} + \lie_{\ggt} {T} -\lie_{\gv+\ggt} \wbackdec{T} \nonumber \\
&\approx & \delta {T} -\lie_{\gv} \wbackdec{T},
\label{change}
\end{eqnarray}
where the indexes have been omitted, and in the last approximation we neglected second order terms.

Note that the final result can be obtained, without loss of generality, by choosing $\ggt^\mu = -\gv^\mu$ from
the beginning, yielding
\begin{align}
\wbackdec{T}_{\nu_1\nu_2\dots\nu_m}^{\quad\mu_1\mu_2\dots\mu_l} \quad &\longrightarrow \quad \wbackdec{T}_{\nu_1\nu_2\dots\nu_m}^{\quad\mu_1\mu_2\dots\mu_l}, \label{def:transf:gauge:pert:tensors}\\
\delta{}T_{\nu_1\nu_2\dots\nu_m}^{\quad\mu_1\mu_2\dots\mu_l} \quad &\longrightarrow \quad \delta{}T_{\nu_1\nu_2\dots\nu_m}^{\quad\mu_1\mu_2\dots\mu_l} - \lie_\gv \wbackdec{T}_{\nu_1\nu_2\dots\nu_m}^{\quad\mu_1\mu_2\dots\mu_l}.\nonumber
\end{align}
This is a very convenient choice given that this specific combination of transformations keeps the functional form of the background tensors the same, and this form should be chosen as the simplest one associated with the preferred coordinate system induced by the underlying symmetry group of the background.

It is this combined transformation (diffeomorphism $\oplus$ coordinate) which is designated as a gauge transformation in the context of perturbation theory in GR. Note that it has different actions on the background and physical geometrical objects, contrary to the usual infinitesimal gauge transformations of GR, and it should also keep the perturbations small. One can identify it as an approximate symmetry group induced by the presence of an extra structure, which is the background manifold and its associated geometrical objects. These objects also induce a preferred coordinate system on the physical manifold, from which one can move just through infinitesimal coordinate transformations. The arbitrariness on the way one introduces this background structure in the physical manifold is at the basis of the gauge dependence of the perturbation problem in GR. Hereafter, a gauge transformation shall mean exactly the above chain of transformations with $\ggt^\mu = -\gv^\mu$.

Given the transformation rule for perturbed tensors Eq.~\eqref{def:transf:gauge:pert:tensors}, the Stewart and Walker (SW) Lemma~\cite{Stewart1974} defines the conditions under which $\lie_\gv \wbackdec{T}_{\nu_1\nu_2\dots\nu_m}^{\quad\mu_1\mu_2\dots\mu_l} = 0$, and hence the conditions for $\delta{}T_{\nu_1\nu_2\dots\nu_m}^{\quad\mu_1\mu_2\dots\mu_l}$ to be GI.  Basically, this Lemma states that first order perturbation of any tensor which is null or a combination of constants with Kronecker deltas in the background is GI. However, within this Lemma there is a hidden assumption that is sometimes overlooked.

The perturbation $\delta{}T_{\nu_1\nu_2\dots\nu_m}^{\quad\mu_1\mu_2\dots\mu_l}$ is defined as the difference $T_{\nu_1\nu_2\dots\nu_m}^{\quad\mu_1\mu_2\dots\mu_l}- \wbackdec{T}_{\nu_1\nu_2\dots\nu_m}^{\quad\mu_1\mu_2\dots\mu_l}$. The conditions in the SW Lemma are sufficient only if the tensor $T_{\nu_1\nu_2\dots\nu_m}^{\quad\mu_1\mu_2\dots\mu_l}$ is defined solely in terms of quantities from the physical manifold, i.e., defined independently of $\Mback$ and $\Upsilon_0$. Generically, if in the definition of $T_{\nu_1\nu_2\dots\nu_m}^{\quad\mu_1\mu_2\dots\mu_l}$ we also use any background tensor, then even if the background tensor $\wbackdec{T}_{\nu_1\nu_2\dots\nu_m}^{\quad\mu_1\mu_2\dots\mu_l}$ is a simple constant its perturbation might not be gauge invariant.

This is a very important distinction that deserves a clear terminology. Thus, we propose the following classification. Any tensor that is defined strictly in terms of objects from a single manifold we shall call a pure tensor. Complementarily, a tensor that involves objects from both manifolds in its definition we shall call a mixed tensor, i.e., it mix objects from the fiducial manifold $\Mback$ with objects from the physical spacetime $\M$. By definition, any background tensor is a pure tensor inasmuch as it is defined solely in terms of tensors from $\Mback$ and the diffeomorphism to map it to $\M$.

The extension of this terminology to perturbed tensors is straightforward. A perturbed tensor shall be called a pure perturbation if it is defined as the difference of two pure tensors. Accordingly, a mix perturbation is defined as the difference of a mixed with a pure tensor. In view of the fact that a background tensor is always a pure tensor, the perturbation might be pure or mixed depending on the nature of the physical tensor in its definition. Now it becomes clear that the SW Lemma applies only for pure perturbations and not for mixed ones.

As an example, let us consider a Friedmann-Lema\^\i tre-Robertson-Walker (FLRW) metric as the background metric. Related to this metric, there is a preferred geodesic vector field $\bn^\mu$ which defines the maximally symmetric spatial hyper-surfaces, i.e., the projection of the background metric $\bg_{\mu\nu}$ induces a maximally symmetric metric in the hyper-surfaces (it has six Killing vectors). Therefore, in this situation, it might be useful to also decompose physical tensors in $\M$ by projecting them with respect to this preferred vector field $\bn^\mu$. One of these tensors is the physical metric $\g_{\mu\nu}$. We can define the mixed scalar $\MP \equiv \g_{\mu\nu}\bn^\mu\bn^\nu / 2$, which gives its projection along the integral curves of $\bn^\mu$. This is a global covariant scalar. The background version of this tensor is simply $\bMP \equiv \bg_{\mu\nu}\bn^\mu\bn^\nu / 2 = -1/2$. Thus, the mixed perturbation associated with these projections reads
\begin{equation*}
\A \equiv \MP - \bMP.
\end{equation*}

Note that $\A$ has been defined in a globally covariant manner. Notwithstanding, at this point it is worth to introduce a coordinate system just to make contact with the usual Bardeen approach. Thus, let us define a coordinate system in which $\bn^\mu = \delta^\mu{}_0$. It is easy to see that in this coordinate system $\g_{\mu\nu}\bn^\mu \bn^\nu=\g_{00} $ and we have $$\g_{00} = \bg_{00}+ 2\A = -1 + 2\A\quad \rightarrow \quad \A=\frac12 \dg_{00}.$$

Therefore, the commonly used metric perturbation $\A$ is simply the mixed perturbation associated with $\MP$. This is a typical example of perturbation that violates the SW Lemma. Indeed, even though the background tensor $\bMP$ is a simple constant, the mixed perturbation $\A$ is a globally covariant scalar but it is not GI.

Generically, any coordinate dependent perturbation can be redefined in a global covariant manner as we have done for $\A$. Furthermore, this procedure does not rely on the symmetries of the background manifold $\bg_{\mu\nu}$. Given an arbitrary tensor, say $F_{\mu\nu}$, a coordinate dependent perturbation associated with its zero-zero component reads$$\delta{}f \equiv \delta{}F_{00} = F_{00} - \wbackdec{F}_{00},$$ where $\wbackdec{F}_{\mu\nu}$ is the background tensor associated with $F_{\mu\nu}$. In order to construct the covariant version of the mixed perturbation $\delta{}f$, we need to define the background vector field $\bn^\nu$ whose integral lines coincide with the zero direction of the above coordinate system. Thus, in this coordinate system $\bn^\nu=\delta^\nu{}_0$. By using this vector field we can define two scalar tensors, namely the mixed $f \equiv F_{\mu\nu}\bn^\mu \bn^\nu$ and the pure $\backdec{f} = \wbackdec{F}_{\mu\nu}\bn^\mu{}\bn^\nu$. Thus, the mixed perturbation is covariantly defined as $\delta{}f = f - \backdec{f}$. Evidently, this is the simplest example as how to define in a covariant manner originally coordinate dependent perturbations. There are more complicated mixed perturbations that we shall analyze in the next section.

Any tensor, independent of its nature as a mixed or pure tensor, transforms as usually under a coordinate transformation. However, they drastically differ under a change of diffeomorphism. While background pure tensors remain unchanged, the mixed tensors can have a very complicated transformation rule depending on its own definition in terms of the combination of tensors from the physical and background manifold. In the case of the above mixed perturbation $\A$, it is easy to find its transformation under a change of diffeomorphism. For that, let us decompose the vector field defining the diffeomorphism as\footnote{All the following definitions are detailed in Section~II of VFP.} $$\gv^\mu = \gvpa\bn^\mu + \gvpe^\mu,\quad \gvpe^\mu = \bhp{\gv^\mu},$$ with $\bhp{\gv^\mu}$ being the projection with respect to $\bh_{\mu\nu}=\bg_{\mu\nu}+\bn_\mu \bn_\nu$. Assuming that $\bn_\mu$ is geodesic, i.e., $\bac_\mu \equiv \bcd_\bn\bn_\mu = 0$ and defining the notation $\dot{T} \equiv \bhp{\lie_\bn{}T}$, it is straightforward that under a gauge transformation, i.e., under a combined coordinate and diffeomorphism transformation with $\ggt^\mu = -\gv^\mu$, we have $$\MP \rightarrow \frac{\bn^\mu\bn^\nu\left(\g_{\mu\nu} - \lie_\gv\bg_{\mu\nu}\right)}{2} = \MP + \dot{\gvpa},$$ and hence
\begin{equation}\label{def:transf:A}
\A \rightarrow \A + \dot{\gvpa},
\end{equation}
as expected. In the above expression the only hypothesis made is that the background foliation is geodesic. Apart from this, Eq.~\eqref{def:transf:A} gives the general transformation for $\A$ in an arbitrary background. However, the other metric perturbations are much more complicated and hence from, here on, we shall restrict ourselves to a FLRW background. Therefore, the background is assumed to be described by:\footnote{All details on our notation and definitions can be found at Section~II of VFP.}
\begin{align}\label{eq:FLRW:SR}
\bEC_{\mu\nu} &= \frac{\bEX}{3}\bh_{\mu\nu}, \qquad\bSR_{\mu\nu} = 2\bK\bh_{\mu\nu}, \\
\bscd_\mu \bEX &= 0 = \bscd_\mu\bK,
\end{align}
with $\bEC_{\mu\nu}$, $\bEX$ and $\bSR_{\mu\nu}$ being respectively the extrinsic curvature, expansion factor and the spatial Ricci tensor, and $\bscd_\mu$ is the spatial covariant derivative of the background. The function $\bK$ is simply a constant divided by the square of the scale factor.

In the next section, we will show how to define in a covariant manner all the commonly coordinate dependent cosmological perturbations characteristic of the Bardeen approach. In addition, using our terminology, we shall compare the covariant version of Bardeen formulation with the usually called covariant approach and elucidate the relation between gauge transformation and change in the slicing of the physical spacetime.

%
\section{Covariant form of Cosmological Perturbations: Bardeen X Covariant approach}
\label{sec:mixten}

Consider a physical spacetime with metric $\g_{\mu\nu}$ and a background metric $\bg_{\mu\nu}$ that we assume to be the FLRW metric. We can always define their difference $\dg_{\mu\nu}$ and decompose it with respect to the FLRW foliation in such a way that
\begin{equation}\label{eq:def:dg}
\dg_{\mu\nu} = 2\A\bn_\mu\bn_\nu + 2\B_{(\mu}\bn_{\nu)} + 2\C_{\mu\nu},
\end{equation}
where
\begin{align*}
\A \equiv \frac12 \dg_{\bn \bn}, \quad \B_\mu \equiv - \bhp{\dg_{\bn\mu}} ,\quad \C_{\mu\nu} \equiv \frac12 \bhp{\dg_{\mu\nu}}.
\end{align*}

The above notation is such that a $\bn$ index means projections in the vector field $\bn^\mu$, i.e., for instance $\dg_{\bn\mu}=\dg_{\alpha\mu}\bn^\alpha$. The above three tensors are mixed tensors. The definition of  $\A$ follows the same line of reasoning as before while the $\B_\mu$ and $\C_{\mu\nu}$ can be define in a covariant manner through the four tensors
\begin{align*}
\MP_\mu &= \bhp{\g_{\mu\bn}}, \quad \bMP_\mu = \bhp{\bg_{\mu\bn}} = 0,\\
\MP_{\mu\nu} &= \frac{\bhp{\g_{\mu\nu}}}{2}, \quad \bMP_{\mu\nu} = \frac{\bhp{\bg_{\mu\nu}}}{2} = \frac{\bh_{\mu\nu}}{2},
\end{align*}
such that $$\B_\mu = \MP_\mu-\bMP_\mu, \quad \C_{\mu\nu} = \MP_{\mu\nu}-\bMP_{\mu\nu}.$$

Note that the above definitions are general and do not depend on any assumption about the smallness of these tensors. In other words, we can always introduce a tensor $\bg_{\mu\nu}$ and define the difference $\dg_{\mu\nu} \equiv \g_{\mu\nu} - \bg_{\mu\nu}$ without assuming anything about $\dg_{\mu\nu}$. Additionally, given a global foliation defined by a timelike vector field $\bn^\mu$ ($\bn^\mu\bn^\nu\bg_{\mu\nu} = -1$) we can define the projector $\bh_{\mu\nu} \equiv \bg_{\mu\nu} + \bn_\mu\bn_\nu$. Therefore, in principle, we can always rewrite Einstein's equations in terms of $\A$, $\B_\mu$ and $\C_{\mu\nu}$ and obtain non-linear second order equations of motion for $\dg_{\mu\nu}$ that encode the same information as those written in terms of $\g_{\mu\nu}$. In this sense, without making the perturbative hypothesis, we are just using different variables to describe the metric $\g_{\mu\nu}$ in terms of the decomposition made in Eq.~\eqref{eq:def:dg} and the given metric $\bg_{\mu\nu}$. This decomposition also shows that the fields $\A$, $\B_\mu$ and $\C_{\mu\nu}$ are just the difference $\dg_{\mu\nu}$ projected in the $3+1$ background splitting. Notwithstanding, in the standard model scenario, cosmological perturbations are first order metric perturbations with respect to the FLRW metric. Thus, we shall assume that  $\A$, $\B_\mu$ and $\C_{\mu\nu}$ are first order perturbations and, unless explicitly stated, all objects will have theirs indexes raised and lowered by the background metric.

Using explicitly their definitions we can calculate the gauge transformation for each one of them. For that it is convenient to decompose the perturbations in terms of the scalar, vector and tensor (SVT) decomposition, i.e.,
\begin{align}
\B_\mu &= \bscd_\mu \BS + \BV_{\mu},\\
\C_{\mu\nu} &= \CS\bh_{\mu\nu} - \bscd_\mu \bscd_\nu \CSD + \bscd_{(\nu}\CV_{\mu)} + \CTD_{\mu\nu},
\end{align}
where $\bscd_\mu\BV^{\mu} = \bscd_\mu \CV^\mu{} = \bscd_\mu\CTD^\mu{}_\nu = \CTD_{\mu}{}^{\mu} = 0$ (see Appendix~\ref{sec:SVT} for details). Using the expressions for the kinematic variables in FLRW spacetime and the projections of $\bcd_\nu\gv_{\mu}$ it can be shown that the gauge transformation for the perturbations above are
\begin{align}\label{eq:gt:A}
\A &\rightarrow \A + \dot{\gvpa}, \\ \label{eq:gt:BS}
\BS &\rightarrow \BS + \dot{\gvpe}^\text{s} - \frac{2\bEX}{3}{\gvpe}^\text{s} - \gvpa, \\ \label{eq:gt:CSD}
\CSD &\rightarrow \CSD + {\gvpe}^\text{s}, \\ \label{eq:gt:CS}
\CS &\rightarrow \CS - \gvpa\frac{\bEX}{3},
\end{align}
where we have decomposed $\gvpe_\mu$ which is the spatial part of the vector field $\gv^\mu$ as $\gvpe_\mu = \bscd_\mu\gvpe^\text{s}+\gvpe_\mu^\text{v}$ [see Eq.~\eqref{eq:dec:vec}]. For the vector sector we find
\begin{align}
\BV^\mu &\rightarrow \BV^\mu + \dot{\gvpe}^{\text{v}\mu}, \\
\CV^\mu &\rightarrow \CV^\mu - \gvpe^{\text{v}\mu}.
\end{align}

Apart from the background slicing, in many cases, it is useful to define a $3+1$ splitting also in the physical manifold. Thus, we introduce an arbitrary global timelike vector field $\n^\mu$. In principle this splitting is completely arbitrary, however, since we are interested in foliations ``close'' to the background slicing, we assume that $\n^\mu$ is such that $\dn_\mu \equiv \n_\mu - \bn_\mu$ is of the same order of $\dg_{\mu\nu}$.
At first order the inverse metric perturbation is $\dg^{\mu\nu} = -\dg_{\alpha\beta}\bg^{\alpha\mu}\bg^{\beta\nu}$. Thus, the normalization of  $\n_\mu$ requires that $$\n_\mu\n_\nu\g^{\mu\nu} = -1 + 2\dn_\mu\bn^\mu - 2\A = -1,\quad \Rightarrow\quad \dn_\mu\bn^\mu = \A.$$

This result can be expressed as
\begin{equation}\label{eq:dn}
\dn_\mu = -\A\bn_\mu + \sn_\mu,
\end{equation}
where we have defined the spatial projection of $\dn_\mu$ as $\sn_\mu \equiv \bhp{\dn_\mu}$.\footnote{It is more convenient to define the normal vector perturbation using its covariant form. Note that, by doing so, we have, at first order, $$\dn^\mu = \bn_\mu\dg^{\mu\nu} + \dn_\mu\bg^{\mu\nu} = 2\A\bn^\nu + \B^\nu + \dn_\mu\bg^{\mu\nu}.$$ Hence the contravariant version of its spatial projection will involve metric perturbations $\B^\mu$ explicitly, i.e., $\bhp{\dn^\mu} = \B^\mu + \sn^\mu$.} Equation~\eqref{eq:dn} shows us that it is the perturbation $\sn_\mu$ that parametrize the freedom in the choice of the spatial hyper-surfaces in $\M$.

Even though this choice of spatial hyper-surface in $\M$ is arbitrary, one could argue that it would be reasonable to use the background slicing itself to foliate the physical space-time, i.e., to use the same vector field $\bn^\mu$ to define the foliation in $\M$. To use this vector field we must first normalize it with respect to the metric $\g_{\mu\nu}$, which gives $$\frac{\bn_\mu}{\sqrt{-\bn_\alpha\bn_\beta\g^{\alpha\beta}}} \approx (1 - \A)\bn_\mu.$$

One can easily see that this choice of spatial sectioning coincide with choosing $\sn_\mu = 0$ in Eq.~\eqref{eq:dn}. Then, in general, we can use the perturbation defined in Eq.~\eqref{eq:dn} and set $\sn_\mu = 0$ to obtain the results in terms of the background foliation.

Notice that by choosing $\sn_\mu = 0$ we are in fact changing the nature of the perturbations as being mixed or pure tensors. Indeed, by attaching the physical foliation with respect to the background slicing, any pure tensor in $\M$ that depends on the vector field $\n^\mu$ automatically becomes a mixed tensor since $\n^\mu$ itself becomes dependent on background quantities. Thus, perturbations with $\sn_\mu = 0$ do not satisfy the SW Lemma. On the other hand, it becomes clear that pure perturbations has one extra variable with respect to mixed perturbations, which is precisely the quantity $\sn_\mu$. In addition, keeping $\sn_\mu$ arbitrary makes the perturbations to satisfy the SW Lemma.

We can further decompose the $\sn_\mu$ [Eq.~\eqref{eq:dec:vec}] as
\begin{equation}\label{eq:sn:dec}
\sn_\mu = \bscd_\mu\sns + \snv_\mu.
\end{equation}

Recalling its definition, i.e., $\sn_\mu \equiv \bhp{\n_\mu} - \bhp{\bn_\mu} = \bhp{\n_\mu}$, we see that under a gauge perturbation this object transforms as
\begin{align}\nonumber
\sn_\mu &\rightarrow \bhp{\n_\mu - \lie_\gv\bn_\mu} = \sn_\mu + \bscd_\mu\gvpa, \\ \label{eq:gt:vms}
\sns &\rightarrow \sns + \gvpa,\qquad \snv_\mu\rightarrow\snv_\mu.
\end{align}

Equations~(\ref{eq:gt:A}--\ref{eq:gt:vms}) allow us to find the transformation rule for all kinematic variables. The first one we shall consider is the acceleration of the curves defined by $\n^\mu$. In what follows, we will construct the pure and mixed perturbations for all kinematics variables but we need to distinguish them without having to change their symbols. Thus, we shall introduce a small circle above each tensor if they are pure tensors and maintain the mixed tensors without any symbol. From Eq.~(A7) of VFP, the pure perturbation of the acceleration is
\begin{equation}\label{eq:def:dac}
\puredac_\mu = \bscd_\mu \dot{\sns} + \dot{\snv}_\mu - \bscd_\mu \A,
\end{equation}
and it is easy to check that $\puredac_\mu$ is GI, which is expected from SW Lemma. If, however, we had used the mixed perturbations ($\sn_\mu = 0$), $\dac_\mu$ would no longer be GI, i.e., the curves with tangent $\bn^\mu$ in the perturbed manifold would not be geodesics and its acceleration would depend on the gauge choice.

The Frobenius theorem guarantees that for a global foliation the field $\n_\mu$ must satisfy $\n_{[\mu}\cd_{\nu}\n_{\alpha]} = 0$. In Eq.~(A11) of VFP we have shown that, at first order, this equation reduces to $\bscd_{[\mu}\sn_{\nu]} = 0$, i.e., the vector part of $\sn_\mu$ is null ($\snv_\mu=0$). Notwithstanding, we will maintain $\snv_\mu$, but with the warning that in this case the decomposition is not global. In other words, the commutator of the operators $\scd_\mu$ will not define a global spatial curvature tensor.

The shear tensor can be calculated using the results of Appendix~A of VFP.\footnote{We have changed the notation on the shear decomposition to avoid an excess of indexes. The map of our notation here with the VFP paper is
\begin{align*}
\dSHs \equiv \dSH^\text{s}, \quad \dSHv_\mu \equiv \dSH^\text{v}_\mu.
\end{align*}
} The pure version of the shear perturbation reads
\begin{equation}\label{eq:def:dSH}
\begin{split}
\puredSH_{\mu\nu} &= \bscd_{\la\mu}\bscd_{\nu\ra}\left(\dSHs + \sns\right) \\
&+ \bscd_{(\mu}\dSHv_{\nu)} + \bscd_{(\mu}\snv_{\nu)} + \dot{\CTD}_\mu{}^\alpha\bh_{\alpha\nu}, \\
\dSHs &\equiv \left(\BS-\dot{\CSD} + \frac{2}{3}\bEX\CSD\right),\quad \dSHv^\mu \equiv \BV^\mu + \dot{\CV}^\mu,
\end{split}
\end{equation}
where the symbol $\la\ra$ represents the projection to the symmetric traceless part of a tensor [Eq.~\eqref{eq:def:sstl}]. One can readily obtain that when performing a gauge transformation, $$\dSHs \rightarrow \dSHs - \gvpa,$$ so that the quantity $\dSHs + \sns$ is GI, which again is a result of the SW Lemma. Once more, for its mixed version, the scalar part of the shear perturbation becomes $\dSH_{\mu\nu} = \bscd_{\la\mu}\bscd_{\nu\ra} \dSHs $, which is not GI.

Besides the shear we need the expansion factor $\EX$ to describe the extrinsic curvature. This field when defined as a mixed perturbation gives,
\begin{equation}
\dEX = \blap\dSHs + \bEX\A + 3\dot{\CS},
\end{equation}
or when considered as a pure perturbation changes to $$\puredEX = \dEX + \blap\sns.$$

It is worth to stress that for the expansion perturbation both forms are gauge dependent since its background value is non-trivial but the pure expansion perturbation is also foliation dependent.

To close the set of kinetic variables we calculate the perturbations of the traceless spatial Ricci tensor
\begin{equation}
\SRtl_{\mu\nu} \equiv \SR_{\la\mu\nu\ra}, \quad \SR_{(\mu\nu)} = \SRtl_{\mu\nu} + \frac{\SR\h_{\mu\nu}}{3},
\end{equation}
which when defined through a pure tensor is naturally GI. Its pure perturbation is given by
\begin{equation}\label{eq:def:dSR}
\begin{split}
\puredSRtl_{\mu\nu} = &-\bscd_{\la\mu}\bscd_{\nu\ra}\left(\CS + \frac{\bEX}{3}\sns\right) \\&
- \frac{\bEX}{3}\bscd_{(\mu}\snv_{\nu)} - (\bscd^2-2\bK)\CTD_{\mu\nu}.
\end{split}
\end{equation}

Once more we have the same situation. For an arbitrary foliation the perturbation for the traceless spatial Ricci is GI but with the caveat that we have introduced an arbitrary field $\sn_\mu$.

Apart from tensor perturbations, it is also important to extend this formalism for scalar quantities. Since the FLRW metric is homogenous and isotropic, one can construct pure perturbations that will be GI by taking spatial gradients of the background fields. Consider for instance the scalar field $\mf$. Its perturbations under a gauge transformation change as
\begin{equation*}
\dmf \rightarrow \dmf - \lie_\gv\bmf = \dmf - \gvpa\dot{\bmf},
\end{equation*}
where we are assuming that the background version of $\mf$ is homogeneous in the hyper-surfaces. However, the gradient $\scd_\mu\mf$ at first order is expressed as
\begin{equation}\label{eq:def:gen:gi}
\scd_\mu\mf \approx \sn_\mu\dot{\bmf} + \bscd_{\mu}\dmf = \bscd_{\mu}\left(\dmf + \sns\dot{\bmf}\right) + \dot{\bmf}\snv_\mu
\end{equation}

Since $\bscd_\mu\bmf=0$, we have that $\delta( \scd_\mu\mf)=\scd_\mu\mf$. Note that the particular combination $\dmf + \sns\dot{\bmf}$ is GI. Furthermore, this pure perturbation compare spatial gradients defined in different spatial sections that causes the appearance of the foliation dependent field $\sn_\mu$.

This is another almost general rule, i.e., GI tensors constructed in the usual covariant approach will depend on the choice of spatial foliation. There are some few exceptions to this rule, such as the Weyl tensor or, at first order, its projections that defines its electric and magnetic parts.

In the case of scalar fields, it is possible to build gauge and foliation invariant tensors. Defining the tensor $$\mfIt_{\mu\nu} \equiv \scd_{\la\mu}\scd_{\nu\ra}\mf - (\lie_\n\mf)\SH_{\mu\nu}$$ we obtain
\begin{equation}
\dmfIt_{\mu\nu} = \bscd_{\la\mu}\bscd_{\nu\ra}\left(\dmf - \dot{\bmf}\dSHs\right).
\end{equation}

Note that the above perturbation is the derivative of a GI quantity that it is usually defined in the so called coordinate approach as the GI scalar perturbation associated with $\mf$. Notwithstanding, we stress once again that all the above perturbed quantities were defined in a covariant manner and with no reference to any coordinate system.

To complete the set of variables, we consider the energy momentum tensor projections
\begin{equation}
\T_{\mu\nu} = \pureed\n_\mu\n_\nu + 2\n_{(\mu}\purenm_{\nu)} + \purep\h_{\mu\nu} + \pureAP_{\mu\nu},
\end{equation}
where $\purenm_\mu \equiv -\hp{\T_\mu{}^\n}$ is the fluid flow of the energy momentum tensor, $\pureed \equiv T_\mu{}^\nu\n_\nu\n^\mu$ the energy density, $\purep \equiv \frac13T_\mu{}^\nu \h_\nu{}^\mu$ the isotropic pressure and $\pureAP_{\mu\nu} \equiv \hp{T_{\mu\nu}}$ the anisotropic pressure spatial traceless tensor. All of them are defined with respect to the foliation defined by $\n^\mu$.

We can define the gradients of the background quantities (see~\cite{Ellis1989}) as
\begin{equation}
\ged_\mu = \kp\scd_\mu\pureed, \qquad \gp_\mu = \kp\scd_\mu\purep, \qquad \gEX_\mu \equiv \scd_\mu\pureEX,
\end{equation}
where we have added the gradient of the expansion factor for completeness. Considering pure perturbations these gradients will be GI and, using Eq.~\eqref{eq:def:gen:gi}, we obtain at first order
\begin{equation}
\begin{split}
&\delta\ged_\mu = \kp\bscd_{\mu}(\ded +  \sns\dot{\bed}) , \quad \delta\gp_\mu = \kp \bscd_{\mu} (\dpp +  \sns\dot{\bp}), \\
&\delta{}\gEX_\mu = \bscd_{\mu} \puredgEX,
\end{split}
\end{equation}
where we have defined the following mixed perturbations
\begin{align}
\ded \equiv \ed - \bed \approx \pureded \qquad \dpp \equiv \p - \bp \approx \puredp,
\end{align}
with
\begin{equation}
\ed \equiv T_\mu{}^\nu\bn_\nu\bn^\mu,\quad\p \equiv \frac13T_\mu{}^\nu \bh_\nu{}^\mu,\quad\pureed \approx \ed,\quad\purep \approx \p,
\end{equation}
and the field
\begin{equation}
\puredgEX = \puredEX + \sns\dot{\bEX}.
\end{equation}

The vector field $\nm_\mu$ when defined as a pure perturbation will also be GI, it is easy to show that
\begin{equation}
\begin{split}
\purednm_\mu &\approx (\bed + \bp) (\snm_\mu - \sn_\mu), \\
&\approx (\bed + \bp) \left[\bscd_{\mu}\left(\snms - \sns\right)+\snmv_\mu-\snv_\mu\right],
\end{split}
\end{equation}
where $$\snm_\mu \equiv -\frac{\bhp{\T_{\mu}{}^{\nu}\bn_\nu}}{\ed+\p}, \quad \snm_\mu = \bscd_{\mu} \snms + \snmv_\mu, \quad \bscd_{\mu} \snmv^\mu{} = 0.$$

This vector field has a simple interpretation. The quantity $(1-\A)\bn_\mu + \snm_\mu$ is, at first order, a timelike eigenvector of $\T_\mu{}^\nu$. Thus, in the fluid frame, $\nm_\mu = 0$ and $\sn_\mu = \snm_\mu$.

Once more we identify the explicit dependency of pure perturbations on the choice of hyper-surface. It also shows that it is the additional degree of freedom introduced by an arbitrary choice of hyper-surface in $\M$ that makes the variables GI.

When working with pure perturbations, one usually chooses the hyper-surface using a physical criteria. For example, one can choose a frame in which there is no particle or energy flux (see~\cite{Ellis1989,Bruni1992}). In the latter case, the field $\sn_\mu = \snm_\mu$ describes a physical property of the system which fixes the extra variable associated with the physical foliation in $\M$.

It is possible to mimic this scenario using mixed tensors and gauge dependent variables. Initially we have the situation where these variables are defined in the background induced frame, which, at this point, is arbitrary. We also have the projected physical hyper-surface vector field discussed above, which is a mixed tensor, described by Eq.~\eqref{eq:sn:dec}, i.e., $\sn_\mu \equiv \bhp{\n_\mu}$. Now, performing a gauge transformation all these variables changes, but it also changes the background induced hyper-surface where they are defined. In Table~\ref{tab:cmp:pure:mixed} we compare the expressions of the kinematic and matter variables.\footnote{Note that in Table~\ref{tab:cmp:pure:mixed} we used a more compact notation for the background spatial derivative, i.e., $\bscd_\mu\equiv {}_{\scp\mu}$, as defined in VFP.}

The first important point is that pure tensors have in general (for the vector and tensor quantities) an additional vector mode. This happens because the background normal vector field is hyper-surface orthogonal but $\n_\mu$ is not, i.e., it does not necessarily define a global foliation.

Let us for instance consider the case when $\n_\mu$ is indeed hyper-surface orthogonal. In this case $\snv_\mu = 0$ and any choice of hyper-surface in $\M$ can be reproduced by a gauge choice $\gvpa = -\sns$. There are two hyper-surfaces defined in $\M$, the induced background $\bSM$ and the physically defined $\SM$. Performing a gauge transformation, we change the diffeomorphism between $\Mback$ and $\M$, which also changes the background foliation $\bSM$. Thus, given a physically defined hyper-surface $\SM$ we can always make a gauge transformation to obtain $\bSM = \SM$. Note, however, that even with this choice of gauge the mixed tensor perturbations would not be gauge invariant. Any further gauge transformation would change all quantities and make $\bSM \neq \SM$. On the other hand, if $\n_\mu$ is not hyper-surface orthogonal we can still choose $\gvpa = -\sns$ but then the vector mode variables will differ while the scalar modes remains equal.

\begin{table}
\caption{\label{tab:cmp:pure:mixed}  Pure and mixed tensor comparison.}
\begin{ruledtabular}
\begin{tabular}{l|l}
\\[-0.7em]
Pure tensors & Mixed tensors \\ \\[-0.7em]\hline\\[-0.7em]
$\pureac_\mu \approx \left(\dot{\sns} - \A\right)_{\scp\mu} + \dot{\snv}_\mu$ & $\ac_\mu \approx -\left(\A+\dot{\gvpa}\right)_{\scp\mu}$ \\ \\[-0.7em]\hline\\[-0.7em]
$\begin{aligned} \pureSH_{\mu\nu} &\approx \left(\dSHs + \sns\right)_{\scp\la\mu\nu\ra} + \snv_{(\mu\scp\nu)} \\ &+ \dSHv_{(\mu\scp\nu)} + \dot{\CTD}_\mu{}^\alpha\bh_{\alpha\nu} \end{aligned}$ & $\begin{aligned} \SH_{\mu\nu} &\approx \left(\dSHs-\gvpa\right)_{\scp\la\mu\nu\ra} \\ &+ \dSHv_{(\mu\scp\nu)} + \dot{\CTD}_\mu{}^\alpha\bh_{\alpha\nu} \end{aligned}$ \\ \\[-0.7em]\hline\\[-0.7em]
$\begin{aligned} \pureSRtl_{\mu\nu} &\approx -\left(\CS + \frac{\bEX}{3}\sns\right)_{\scp\la\mu\nu\ra} \\ &- \frac{\bEX}{3}\snv_{(\mu\scp\nu)} \\ &- (\bscd^2-2\bK)\CTD_{\mu\nu} \end{aligned}$ & $\begin{aligned} \SRtl_{\mu\nu} &\approx -\left(\CS - \frac{\bEX}{3}\gvpa\right)_{\scp\la\mu\nu\ra}\\ & - (\bscd^2-2\bK)\CTD_{\mu\nu} \end{aligned}$ \\ \\[-0.7em]\hline\\[-0.7em]
$\pureged_\mu = \kp\left(\ded + \sns\dot{\bed}\right)_{\scp\mu}$ & $\ged_\mu = \kp\left(\ded - \gvpa\dot{\bed}\right)_{\scp\mu}$ \\ \\[-0.7em]\hline\\[-0.7em]
$\puregp_\mu = \kp\left(\dpp + \sns\dot{\bp}\right)_{\scp\mu}$ & $\gp_\mu = \kp\left(\dpp - \gvpa\dot{\bp}\right)_{\scp\mu}$ \\ \\[-0.7em]\hline\\[-0.7em]
$\puregEX_\mu = \left(\dEX + \blap\sns + \sns\dot{\bEX}\right)_{\scp\mu}$ & $\gEX_\mu = \left(\dEX - \blap\gvpa - \gvpa\dot{\bEX}\right)_{\scp\mu}$ \\ \\[-0.7em]\hline\\[-0.7em]
$\begin{aligned} \purenm_\mu &= (\bed + \bp)\left(\snms - \sns\right)_{\scp\mu} \\ &+ (\bed + \bp)(\snmv_\mu-\snv_\mu) \end{aligned}$ & $\begin{aligned} \nm_\mu &= (\bed + \bp)\left(\snms + \gvpa\right)_{\scp\mu} \\ &+ (\bed + \bp)\snmv_\mu \end{aligned}$ \\ \\[-0.7em]\hline\\[-0.7em]
$\hp{\n_\mu} = 0$ & $\bhp{\n_\mu} = \left(\sns + \gvpa\right)_{\scp\mu} + \snv_\mu$
\end{tabular}
\end{ruledtabular}
\end{table}

The above discussion shows that, concerning scalar perturbations, the freedom in choosing a specific foliation in the GI covariant formalism is equivalent to choose a gauge in the Bardeen approach. Accordingly, both situations suffer from the same difficulties. For instance, in a geodesic foliation we have $\ac_\mu \approx \bscd_\mu(\dot{\sns} - \A) = 0$, but this condition does not fix completely the hyper-surfaces. For any other choice of $\sns + f$ such that $\dot{f} = 0$ we would still have $\dac_\mu = 0$. This is exactly the problem with the synchronous gauge, which generates the same unphysical modes in the solutions.

Concluding, it is equivalent to work with the GI covariant formalism with a fixed choice of foliation or with the gauge dependent formalism with a fixed gauge. The only advantage of the former was the maintenance of its covariance. However, since we have constructed a fully covariant formalism to describe the metric perturbations, both methods become equivalent. Besides, when one needs to go beyond a Friedmann background metric, it is not clear what tensors are GI. Hence, for a general background, our method proves to be straightforward in obtaining covariant perturbations.

\section{Gauge and Foliation Invariant Variables}
\label{sec:gi}

In the last section we have shown that pure or mixed perturbations are respectively foliation and gauge dependent. This analysis was based in first order expansion of kinematic quantities. In this section we shall construct exact tensors that in first order reduce to the usual Bardeen's potentials, i.e., reduce to $\AI$ and $\CSI$. There are two possible ways: we can start with mixed tensors which are gauge dependent and combine them to obtain first order GI quantities or we can begin with pure tensors which are already GI and look for foliation independent combinations. Both methods arrive at the same variables which means that Bardeen GI variables are not just GI but are also independent of the choice of the foliation.

Let us consider first the potential $\CSI$. Its common definition is
\begin{equation}\label{def:CSI}
\CSI \equiv \CS - \frac{\bEX}{3}\dSHs.
\end{equation}

Looking at Table~\ref{tab:cmp:pure:mixed} we immediately see that $\CS$ appears in $\SRtl_{\mu\nu}$ and $\dSHs$ in $\SH_{\mu\nu}$. Thus, an obvious choice is
\begin{equation}\label{eq:def:CSIt}
\CSIt_{\mu\nu} = - \SRtl_{\mu\nu} -\frac{\EX}{3}\SH_{\mu\nu},
\end{equation}

It is straightforward to calculate that at first order $\CSIt_{\mu\nu}$ reads
\begin{equation}
\begin{split}
\dCSIt_{\mu\nu} &= \bscd_{\la\mu}\bscd_{\nu\ra}\CSI + (\bscd^2-2\bK)\CTD_{\mu\nu} \\
&-\frac{\bEX}{3}\left(\bscd_{(\mu}\dSHv_{\nu)} + \dot{\CTD}_\mu{}^\alpha\bh_{\alpha\nu}\right).
\end{split}
\end{equation}

Therefore, at first order, $\CSIt_{\mu\nu}$ depends only on gauge invariant variables and its scalar mode is just the Bardeen potential $\CSI$. Note also that using mixed or pure tensors results in the same expression for $\dCSIt_{\mu\nu}$, which means that this variable is GI and foliation invariant.

To define the next Bardeen potential $\AI$ defined by
\begin{equation}\label{def:AI}
\AI \equiv \A + \dot{\dSHs},
\end{equation}
we introduce the following tensor
\begin{equation}
\begin{split}
\Ac_{\mu\nu} &\equiv \scd_{\la\mu}\ac_{\nu\ra} + \ac_{\la\mu}\ac_{\nu\ra}, \\
\dAc_{\mu\nu} &= \bscd_{\la\mu}\bscd_{\nu\ra}\left(\dot{\sns}-\A\right) + \bscd_{\la\nu}\dot{\snv}_{\mu \ra},
\end{split}
\end{equation}
where the term $\ac_{\la\mu}\ac_{\nu\ra}$ was include for later convenience even though being of second order it does not alter the first order perturbation. The desired tensor is
\begin{equation}\label{eq:def:AIt}
\AIt_{\mu\nu} \equiv (\lie_\n\SH)_{\la\mu\nu\ra}-\Ac_{\mu\nu},
\end{equation}
where $(\lie_\n\SH)_{\la\mu\nu\ra}$ means that one has to first take the Lie derivative and then apply the projector. Its perturbation reads
\begin{equation}
\dAIt_{\mu\nu} = \bscd_{\la\mu}\bscd_{\nu\ra}\AI + \bscd_{\nu}\dot{\dSHv}_{\mu} + \ddot{\CTD}_\mu{}^\alpha\bh_{\alpha\nu} + \frac{2\bEX}{3}\dot{\CTD}_\mu{}^\alpha\bh_{\alpha\nu},
\end{equation}
whose scalar mode gives the usual GI potential $\AI$ but defined in  a covariant manner. This tensor share the same properties of $\CSIt_{\mu\nu}$, i.e., at first order it is gauge and foliation independent.

The two tensors defined above in Eqs.~\eqref{eq:def:CSIt} and \eqref{eq:def:AIt}
are closely related to the electric part of the Weyl tensor $\WeylE_{\mu\nu} = \Weyl_{\mu\alpha\nu\beta}\n^\alpha\n^\beta$. In terms of the kinetic variables we have
\begin{equation*}
\begin{split}
\WeylE_{\mu\nu} &= \frac{1}{2}\left(\frac{\EX}{3}\SH_{\mu\nu}-(\lie_\n\SH)_{\la\mu\nu\ra}+\SRtl_{\mu\nu} + \Ac_{\mu\nu}\right), \\
&= -\frac{1}{2}\left(\CSIt_{\mu\nu} + \AIt_{\mu\nu}\right).
\end{split}
\end{equation*}

Thus these gauge and hyper-surface invariant variables naturally appears as parts of the Weyl tensor. In fact, this should have already been expected since the Weyl tensor is null in the Friedmann background and differently from the other examples it does not depend on the choice of foliation in $\M$. Its electrical part however could depend on the foliation but since the background Weyl tensor is null at first order $\WeylE_{\mu\nu}$ will also not depend on the choice of foliation, i.e.,
\begin{equation}
scalar \ part \ of  \left(\delta\WeylE_{\mu\nu}\right) = -\bscd_{\la\mu}\bscd_{\nu\ra}\frac{1}{2}\left(\AI+\CSI\right).
\end{equation}

To obtain the GI perturbation associated with the expansion factor we define the tensor
\begin{equation}\label{fullexpansion}
\begin{split}
\EXIt_{\mu\nu} &\equiv 3(\lie_\n\CSIt)_{\la\mu\nu\ra} + \EX\AIt_{\mu\nu}, \\
&= -3(\lie_\n\SRtl)_{\la\mu\nu\ra}-(\lie_\n\EX)\SH_{\mu\nu} - \Ac_{\mu\nu},
\end{split}
\end{equation}
which at first order gives
\begin{equation}
\dEXIt_{\mu\nu} = \bscd_{\la\mu}\bscd_{\nu\ra}\EXI,\quad \EXI \equiv 3\dot{\CSI} + \bEX\AI = \dEX - \blap\dSHs - \dot{\bEX}\dSHs.
\end{equation}
In the same manner as the previous quantities, this perturbation is also gauge and foliation invariant.

In the above reasoning we have constructed three gauge and foliation invariant variables $\AI$, $\CSI$ and $\EXI$ that are related respectively to the acceleration of $\n^\mu$, the spatial Ricci tensor and expansion factor. These variables were made GI by combining them with the shear tensor.

We should stress that we could have alternatively built  them directly from the mixed perturbations and hence without referring to a second spatial folitation in $\M$. In this case, these variables have simpler meaning when compared with their pure versions. By using the mixed perturbations we are comparing tensors that are defined in the same spatial sectioning but with different metrics.

On the other hand, when dealing with pure perturbations we are comparing tensors defined with respect to the $\n^\mu$ foliation with those defined with respect to the $\bn^\mu$ background foliation. As a consequence, it appears an additional term in these perturbations and the shear perturbation is given by Eq.~\eqref{eq:def:dSH}. The presence of the field $\sns$ comes from the fact that we are comparing tensors in different hyper-surfaces.

In the gravitational sector, there are only two variables that are simultaneously gauge and foliation invariant. The $\AI$ and $\CSI$ potentials are sufficient to describe the evolution of the perturbations and there is no need to deal with the additional degree of freedom coming from $\sn_\mu$. Notwithstanding, the usual covariant approach formulation demands for consistency the inclusion of $\sn_\mu$ which from this point of view seems unnecessary.

In summary, the usual covariant approach introduces an a priori arbitrary foliation described by $\sns$ and uses it to build GI quantities. In contrast, the Bardeen approach uses the shear potential $\dSHs$ to construct the GI quantities and has the advantage of not introducing an additional degree of freedom to the system.

For the matter sector, its own degrees of freedom can be used to defined a physically motivated notion of spatial hyper-surfaces and consequently $\sns$. For example, in the case of a perturbed perfect fluid we can define the vector normal to the spatial sectioning as the time-like eigenvector of the energy momentum tensor which fix $\sns = \snms$ and defines a natural choice for $\sn_\mu$.\footnote{If the fluid it not irrotational this folitation will not be global, which can cause further complications.}  However, it is worth emphasizing that there is nothing special about this choice of foliation. Alternatively, we could as well have defined the spatial sectioning by requiring the GI acceleration vector field $\ac_\mu$ to be null. Thus, even though $\sn_\mu$ could be viewed as dynamically superfluous, its importance relies in the possibility to use physically relevant choices of spatial sectioning.

\section{Comparison between the two methods}

Usually the gauge issue is seen as the result of introducing a background manifold, as an absolute object, to the problem in hand. While this is true, one can trace back what are the principles which led to this construction. First, one states the physical problem by describing it with a metric manifold and a matter content. At this point all the description is invariant under diffeophormisms. The next and key step is to assume that the physical system can be approximate by a fiducial and hopefully simpler model. For example, in the usual covariant approach, one states that some kinetic variables are ``small'' for a given choice of hypersurfaces. So, in this approximation, it is necessary to introduce a foliation and then assume special characteristics for the kinetic tensors. However, their ``smallness'' is for certain hypersurfaces. Clearly, there is no way to guarantee this property for a generic choice of foliation. Notwithstanding, given a initial hypersurface where the smallness assumption is valid, one can always find other hypersurfaces in which the same assumption still holds. These other hypersurfaces can be related to the first one by a diffeomorphism. Such diffeomorphism should be applied only to the hypersurface, to describe the change in the foliation.

More precisely, given only the physical quantities, one introduces the hypersurface choice by defining a normal vector field $n^\mu$. Using this field one obtains all kinetic variables as functions of the metric and their covariant derived objects. For example, the extrinsic curvature is a complicated function of $n^\mu$, i.e., $\EC(n, \g)_{\mu\nu}$. However, by Proposition 8.3.13 of~\cite{Wald1984}, a hypersurface expressed by $n^\mu$ and another one defined by $s^\mu$ are diffeomorphic.\footnote{Here we are assuming that the hypersufaces are smooth Cauchy surfaces.} Hence, the change of hypersurfaces can be implemented by an exponential map $s^\mu = \exp(\lie_A)n^\mu$, which at first order changes $n^\mu$ as $n^\mu \rightarrow n^\mu + \lie_A n^\mu,$ for an arbitrary but small vector field $A^\mu$. Hence, when one changes the hypersurface, all objects in $\EC_{\mu\nu}$ but $n^\mu$ are kept fixed, whereas $n^\mu$ can be seen as being transformed by a diffeomorphism.

This is exactly what happens when one uses the background foliation to build the mixed kinetic variables, $\EC(\bn,\g)_{\mu\nu}$, where $\bn$ is the normal vector field used to define the background foliation. Note that, as discussed in Sec.~\ref{sec:pert:sp}, any change in the mapping of the background quantities into the physical manifold will change all the background quantities, and of course the normal vector $\bn^\mu$, and keep fixed all physical quantities (remember that the subsequent coordinate transformation we introduced to complete what we defined as a gauge transformation was just a matter of convenience in order to reverse this situation). This change in the mapping can also be seen as a diffeomorphism acting only on $\bn^\mu$. Hence this shows that both $\EC(n,\g)_{\mu\nu}$ and $\EC(\bn,\g)_{\mu\nu}$ have exactly the same transformation rule when one changes the spatial sectioning and the gauge, respectively. It is worth emphasizing that the analogy between the gauge and the hypersurface freedom is not only a coincidence at first order but a general result valid at any order and for any background.

Note, however, that gauge transformations and changes of foliation are two physically distinct operations. Their equivalence is restrict only to the situation described above in the cosmological perturbation scenario.

This reasoning show us that one a gauge issue is automatically created when one introduces the approximation hypothesis through a choice of hypersurfaces, i.e., the freedom in choosing the hypersurfaces among all possibilities in which the approximation is valid. 
It is clear that, an arbitrary choice of hypersurface is always possible and this by itself is not connected to the gauge issue. It is only when it is used to implement the perturbative hypothesis that it creates a gauge problem. We thus argue that the gauge freedom is a natural ingredient of the perturbative approach, independently of an eventual explicit introduction of a background manifold.

It is instructive to summarize the discussion above before going further in the higher order issues. The usual covariant analysis can be described as follows:
\begin{enumerate}
\item Assume that, for a given special foliation defined by a vector field $n^\mu$, some kinetic tensors and projected matter tensors are first or zero order.
\item Linearize the equations of motion by removing all second and higher order terms.
\item Use the SW Lemma to obtain gauge invariant (GI) variables. This is done
by manipulating the kinetic variables (applying spatial gradients, removing traces, etc), to define a complete set of gauge invariant tensors and equations of motion to express the physical problem at hand.
\end{enumerate}
The steps above completely remove the gauge ambiguity from the problem. They also avoid the explicit introduction of a background metric and the zero-plus-first-order splitting. Nonetheless, as we can see in Table~\ref{tab:cmp:pure:mixed}, the GI variables used in this context are not foliation invariant. Hence, one must solve this ambiguity by choosing a hypersurface before solving the system. Consequently, the procedure above removes the gauge ambiguity of the problem, but does not address the hypersurface problem.

The coordinate approach of the metric perturbation, on the other hand, follows the steps:
\begin{enumerate}
\item Define a background metric and a matter content within a given coordinate system.
\item Write explicitly the metric and matter perturbations as the difference between the physical objects and their background counterparts.
\item Choose a spatial sectioning in the physical manifold to describe the kinetic variables.
\item Solve the gauge issue by applying one of two different approaches:
\begin{itemize}
\item Make a gauge choice for the perturbations such that the gauge ambiguity is completely (and hopefully) solved.
\item Find combinations between perturbation variables which are gauge invariant.
\end{itemize}
\item Expresses every equation in terms of the perturbations disregarding any second-order terms.
\end{enumerate}
This method starts with a choice of coordinate system adapted to the background metric and, as such, implicitly specifies the background foliation. After introducing a spatial section in the physical manifold, one obtains a set of tensor components describing this hypersurface in a coordinate system adapted to the background foliation. If one chooses to fix the gauge, one should elect some perturbations to have specific values. This choice often involves perturbations related to the choice of hypersurface. Otherwise, one finds the GI combination of perturbations, which again involves (but not necessarily) hypersurface dependent quantities. In the end, the system is given by a combination of choices of the coordinate system, hypersurfaces, and a gauge choice or GI combinations. The high number of different ingredients makes the problem cumbersome, but straightforward to make calculations. This method has potentially the two ambiguities, gauge and foliation, convoluted in the system.

In the two methods described above the spatial sectioning ambiguity is not handled. In both cases, the foliation is defined at the beginning. In the second approach, one may also use the background foliation, as it was done in the review~\cite{Kodama1984}. 

In the present paper, we have outlined how to deal with the two problems as:
\begin{enumerate}
\item Define the background geometry covariantly. In the case of a FLRW metric, one should include the definition of the time-like vector field $\bn^\mu$ characterizing the preferred homogeneous and isotropic foliation.
\item Map the background geometry in the physical manifold and define perturbations as the differences between the physical and the background tensors. Additionally, define mixed tensors by using the background foliation $\bn^\mu$ to express the kinetic variables.
\item Find the mixed tensor combinations which provides the GI variables.
\item Write the equations of motion in terms of these GI variables.
\end{enumerate}

The above procedure provides a coordinate and gauge free approach to the perturbations. Since the gauge freedom and foliation ambiguity have the same transformation rule, as we have shown above (as a mixed tensor combination $Q(\bn,\g)$ transforms under a gauge transformation in the same way as $Q(n,\g)$ transforms under a change in the foliation $n$, then if the mixed tensor combination $Q(\bn,\g)$ is gauge invariant it will also be foliation independent), any GI variable found at any order will be also {\bf foliation invariant}. Therefore, this method provides a straightforward way to obtain gauge and foliation invariant quantities in a covariant fashion.

The higher order approaches for the cosmological perturbations are mostly based on the usual covariant approach. They take advantage of the SW Lemma to define, in a systematic way, higher order GI variables. However, this method does not address the foliation dependency of the perturbations, leading to a system where this ambiguity is resolved like the original gauge problem, i.e., by introducing arbitrary hypersurfaces. For this reason, we argue that the method presented in this paper is more suitable to deal with these issues. It has the advantage of being covariant, independent of any particular choice of foliation, and any GI variable obtained from it will also be foliation independent at any order. However, since the mixed tensors do not satisfy the SW Lemma, to our knowledge there is still no systematic way to obtain GI variables at arbitrary higher order perturbative level.

\section{Conclusions}\label{conclusions}

The purpose of this work was to discuss the cosmological perturbation scenario, and to present some new improvements in the formalism. Usually, the Bardeen approach is viewed as coordinate dependent, even though gauge invariant variables can be defined in the formalism. This coordinate dependence seems to weaken the treatment, specially when one goes to higher order perturbations. Here, as an effort to adequate it, we have presented the Bardeen approach in a complete covariant way.

In the process of building a covariant Bardeen approach, a crucial difference in how to define the perturbed variables appeared. In order to deal with this situation, we proposed a new classification of tensors in terms of the fundamental objects used in their definitions. Pure tensors are formed by objects defined with respect to a single manifold, while mixed tensors combine objects coming from two different manifolds, i.e., they mix background and physical tensors.

The other well known covariant approach defines the cosmological perturbations in a foliation dependent way. The equivalence of the Bardeen and the so called covariant approach has been established in the past but it is common to find authors advocating in favor of one or the other. Therefore, it seemed appropriate to compare both formalism in the light of our new language. We have shown that, apart from their dynamical equivalence, they possess complementary indeterminacies. The Bardeen approach has the usual gauge dependence, while in the covariant formalism it appears a dependence on the space-like hyper-surfaces one uses to perform the space-time foliation. We have shown how these dependencies can be one-to-one mapped to each other. This is evident from inspection of Table~\ref{tab:cmp:pure:mixed}.

The natural question that appears is whether there is a variable that is simultaneously foliation and gauge invariant. The answer is affirmative and, using Table~\ref{tab:cmp:pure:mixed}, we have presented a general way to construct full non-linear tensors that, at first order, are simultaneously foliation and gauge invariant. They are shown in Eqs.~(\ref{eq:def:CSIt},\ref{eq:def:AIt},\ref{fullexpansion}). Their scalar parts correspond exactly to the Bardeen potentials, hence, showing that these potentials are a particular combination that at first order are not only gauge invariant, as it is well known, but also foliation independent.

Hence, the new results of our paper can be summarized as follows:
\begin{enumerate}
\item Formulation of Bardeen's approach in a completely covariant manner
\item A new classification of tensors in perturbation theory: pure and mixed tensors.
\item The construction of table $1$, where the relationship between the indeterminacies of the covariant and Bardeen approaches are clearly shown.
\item The construction of full non-linear tensors that at first order are simultaneously foliation and gauge invariant, exhibited in Eqs.~(\ref{eq:def:CSIt},\ref{eq:def:AIt},\ref{fullexpansion}).
\end{enumerate}

These four results were presented in a logical order where one step cannot be taken without the preceding one.

We expect that the above results can be useful in the examination of more involved problems. For instance, stimulated by the dark energy problem, there has been recently several attempts to construct a robust perturbative formalism to describe metric perturbations. In particular, the possibility of a significant back-reaction contribution due to the formation of non-linear structures in small scales has raised some doubts about the validity of first order perturbations in the standard cosmological model. In order to address these issues, it is crucial to control the gauge and foliation dependency of the perturbed variables. As we will show elsewhere, our formalism is adequate to describe first order perturbations around a FLRW background, and to analyze the necessary conditions for the validity of the first order cosmological perturbations.

\section*{ACKNOWLEDGMENTS}

We would like to thank CNPq of Brazil for financial support. S.D.P.Vitenti acknowledges financial support from Capes, under the program ``Ci\^{e}ncias sem Fronteiras''.

\appendix

\section{Scalar, Vector and Tensor Decomposition}
\label{sec:SVT}

Given a foliation defined by a global timelike vector field $\bn_\mu$ and the spatial covariant derivative $\bscd_\mu$ we define the Laplace-Beltrami operator as $\bscd^2 \equiv \bh^{\mu\nu}\bscd_\mu\bscd_\nu$. Depending on the details of the spatial foliation and the class of functions that we are working with this operator can have a unique inverse (see~\cite{Stewart1990}). Assuming that this is the case, we can decompose any spatial vector field $A_\mu$, first we define $\bscd^2{}A^\text{s} = \bscd_\mu{}A^\mu$, using the unique inverse of $\bscd^2$ we, then, define $A^\text{v}{}_\mu = A_\mu - \bscd_\mu\bscd^{-2}\bscd_\nu{}A^\nu$ such that $\bscd_\mu{}A^\text{v}{}^\mu = 0$ and, therefore,
\begin{equation}\label{eq:dec:vec}
A_\mu = \bscd_\mu{}A^\text{s} + A^\text{v}{}_\mu.
\end{equation}

To decompose a second order spatial tensor we first introduced the spatial symmetric trace-less symbol,
\begin{equation}\label{eq:def:sstl}
T_{\la\mu\nu\ra} \equiv \hp{T_{(\mu\nu)}}-\frac{\h_{\mu\nu}T_{\alpha\beta}\h^{\alpha\beta}}{3}.
\end{equation}
When acting on a spatial second derivative of a scalar field it can be written as
\begin{equation}
A_{\scp\la\mu\nu\ra} = \left(\bscd_\mu\bscd_\nu - \frac{\bh_{\mu\nu}\bscd^2}{3}\right)A,
\end{equation}
where the double divergence of the expression above gives
\begin{equation*}
\bscd_\mu\bscd_\nu A^{\scp\la\mu\nu\ra} = \left(\frac{2\bscd^4}{3} + \bSR^{\mu\nu}\bscd_\mu\bscd_\nu + \bscd^\mu\bSR\bscd_\mu\right)A,
\end{equation*}
for a FLRW background we can use the form of the spatial Ricci tensor [Eq.~\eqref{eq:FLRW:SR}] to write it as
\begin{equation}
\bscd_\mu\bscd_\nu A^{\scp\la\mu\nu\ra} = \frac{2\bscd^2}{3}\left(\bscd^2 + 3\bK\right)A.
\end{equation}
The equation above inspire us to define
\begin{equation}
\blapK \equiv \bscd^2 + \frac{3}{2}\bilap\left(\bSR^{\mu\nu}\bscd_\mu\bscd_\nu + \bscd^\mu\bSR\bscd_\mu\right)
\end{equation}
so that, in general,
\begin{equation}
\bscd_\mu\bscd_\nu A^{\scp\la\mu\nu\ra} = \frac{2}{3}\blap\blapK{}A.
\end{equation}

Assuming now that both $\blap$ and $\blapK$ operators have unique inverse we can write the double divergence of an arbitrary spatial second order tensor $A_{\mu\nu}$ as
\begin{equation}
\bscd_\mu\bscd_\nu{}A^{\la\mu\nu\ra} = \blap\blapK A^\text{s},
\end{equation}
this shows that the tensor
\begin{equation}\label{eq:def:dl:dl}
B_{\mu\nu} = A_{\la\mu\nu\ra} - \bscd_{\la\mu}\bscd_{\nu\ra}\bilapK\bilap\bscd_\alpha\bscd_\beta{}A^{\la\alpha\beta\ra},
\end{equation}
has null double divergence. Then, clearly $\bscd_\mu{}B^\mu{}_\nu$ is a divergenceless spatial vector. To extract this vector we first note that, given a divergenceless spatial vector $F_\mu$ we can write
\begin{equation}
\bscd^\mu\bscd^\nu\bscd_{(\mu}F_{\nu)} = \left(\frac{\bscd_\mu\bSR}{2} + \bSR_{\la\mu\nu\ra}\bscd^\nu\right)F^\mu,
\end{equation}
note that, in general, the double divergence of $\bscd_{(\mu}F_{\nu)}$ is not null, however, when the background is FLRW we have $\bSR_{\la\mu\nu\ra} = 0 = \bscd_\mu\bSR$ this quantity is null. Using Eq.~\eqref{eq:def:dl:dl} we define the tensor
\begin{equation}
H_{\mu\nu} \equiv \bscd_{(\mu}F_{\nu)} - \bscd_{\la\mu}\bscd_{\nu\ra}F^\text{ds},
\end{equation}
where we defined the scalar built above as
\begin{equation}
F^\text{ds} \equiv \bilapK\bilap\bscd^\mu\bscd^\nu\bscd_{(\mu}F_{\nu)}.
\end{equation}
Taking the single divergence of the tensor above we obtain
\begin{equation}
\bscd^\mu{}H_{\mu\nu} = \blapR_\mu{}^\nu F_\nu,
\end{equation}
where we defined the operator
\begin{equation}
\begin{split}
&\blapR_\mu{}^\nu \equiv \frac{\bh_\mu{}^\nu\blap + \bSR_\mu{}^\nu}{2} \\
&+ \left(\frac{2\bscd_\nu\blap}{3}+\bSR_\nu{}^\mu\bscd_\mu\right)\bilapK\bilap\left(\frac{\bscd_\mu\bSR}{2} + \bSR_{\la\mu\nu\ra}\bscd^\nu\right).
\end{split}
\end{equation}
Finally, if the operator $\blapR_\mu{}^\nu$ has a unique inverse we define the tensor
\begin{equation}
E_{\mu\nu} = B_{\mu\nu} - H_{\mu\nu},
\end{equation}
which the divergence is given by  $$\bscd_\mu{}E^\mu{}_\nu = \bscd_\mu{}B^\mu{}_\nu - \blapR_\mu{}^\nu F_\nu,$$ using the unique inverse we choose $F_\nu = \bilapR_\nu{}^\beta\bscd_\alpha{}B^\alpha{}_\beta$, which results in $\bscd_\mu{}E^\mu{}_\nu = 0$.

When we impose that the background is FLRW the operator $\blapR_\mu{}^\nu$ is simply
\begin{equation}
\blapR_\mu{}^\nu = \frac{\bh_\mu{}^\nu}{2}(\blap + 2\bK),
\end{equation}
and the second order tensor $H_{\mu\nu} = \bscd_{(\mu}F_{\nu)}$.

Thus, in general we can decompose a second order tensor $A_{\la\mu\nu\ra}$ as
\begin{equation}\label{eq:dec:ten}
A_{\la\mu\nu\ra} = \bscd_{\la\mu}\bscd_{\nu\ra}(A^\text{s}+F^\text{ds}) + \bscd_{(\mu}F_{\nu)} + E_{\mu\nu},
\end{equation}
where $\bscd_\mu{}E^\mu{}_\nu = 0 = \bscd_\mu{}F^\mu$.



\end{document}